\documentclass[preprint,prc,showpacs,preprintnumbers,
               superscriptaddress,amsmath,amssymb,floatfix]{revtex4}
\usepackage{graphicx,latexsym,amssymb}
\usepackage{multirow,amsmath}

\newcommand{\ben}{\begin{enumerate}}
\newcommand{\een}{\end{enumerate}}
\newcommand{\beq}{\begin{equation}}
\newcommand{\eeq}{\end{equation}}
\newcommand{\beqn}{\begin{eqnarray}}
\newcommand{\eeqn}{\end{eqnarray}}

\newcommand{\beqd}{\begin{eqnarray*}}
\newcommand{\eeqd}{\end{eqnarray*}}
\newcommand{\bea}{\begin{array}}
\newcommand{\eea}{\end{array}}
\newcommand{\bcen}{\begin{center}}
\newcommand{\ecen}{\end{center}}
\newcommand{\btab}{\begin{tabular}}
\newcommand{\etab}{\end{tabular}}
\newcommand{\bsub}{\begin{subequations}}
\newcommand{\esub}{\end{subequations}}
\begin{document}
%\begin{CJK*}{GBK}{song}

\title{
Octupole degree of freedom for the critical-point candidate nucleus $^{152}$Sm in a
reflection-asymmetric relativistic mean-field approach
}

\author{W. Zhang}
\affiliation{State Key Laboratory of Nuclear Physics and Technology, School of Physics,
Peking University, Beijing 100871, People's Republic of China }
\affiliation{School of Electrical Engineering and Automation,
He'nan Polytechnic University, Jiaozuo 454003, People's Republic of China }
\author{Z. P. Li }
\affiliation{State Key Laboratory of Nuclear Physics and Technology, School of Physics,
Peking University, Beijing 100871, People's Republic of China }
\author{S. Q. Zhang \footnote{Email: sqzhang@pku.edu.cn}}
\affiliation{State Key Laboratory of Nuclear Physics and Technology, School of Physics,
Peking University, Beijing 100871, People's Republic of China }
\author{J. Meng \footnote{Email: mengj@pku.edu.cn}}
\affiliation{School of Physics and Nuclear Energy Engineering,
Beihang University, Beijing 100191, People's Republic of China }
\affiliation{State Key Laboratory of Nuclear Physics and Technology,
School of Physics, Peking University, Beijing 100871, People's
Republic of China } \affiliation{Department of Physics, University
of Stellenbosch, Stellenbosch, South Africa}

\date{\today}

\begin{abstract}
The potential energy surfaces of even-even $^{146-156}$Sm are
investigated in the constrained reflection-asymmetric relativistic
mean-field approach with parameter set PK1. It is shown that the
critical-point candidate nucleus $^{152}$Sm marks the shape/phase
transition not only from $U(5)$ to $SU(3)$ symmetry, but also from
the octupole-deformed ground state in $^{150}$Sm  to
the quadrupole-deformed ground state in $^{154}$Sm. By including the octupole
degree of freedom, an energy gap near the Fermi surface for
single-particle levels in $^{152}$Sm with $\beta_2 = 0.14 \sim 0.26$
is found, and the important role of the octupole deformation driving
pair $\nu 2f_{7/2}$ and $\nu 1i_{13/2}$ is demonstrated.
\end{abstract}

% 21.10.-k Properties of nuclei; nuclear energy levels properties
% 21.10.Dr Binding energies and masses
% 21.10.Pc Single-particle levels and strength functions
% 21.10.Gv Nuclear deformation ¡ªnucleon distribution
% 21.60.Jz Nuclear Density Functional Theory and extensions (includes Hartree¨CFock and random-phase approximations)
% 27.60.+j 90<=A<=149
% 27.70.+q 150<=A<=189
\maketitle

\section{Introduction}

The first-order phase transition between spherical $U(5)$ and
axially deformed $SU(3)$ shapes~\cite{Iachello1998,Iachello2001} has
received widespread attention in the past decade. It was shown that
$^{152}$Sm and other $N=90$ isotones are the empirical examples of
the analytic description of nuclei at the critical point of such a
transition~\cite{Casten2001}. Theoretical studies on the phase
transition were typically based on phenomenological geometric
models of nuclear shapes and potentials~\cite{Iachello2001} and
algebraic models of nuclear structure~\cite{Cejnar2009}. The first
calculations, establishing a link between dynamical symmetry models
and microscopic theories, were carried out using the relativistic
mean-field (RMF) approximation in the Sm isotopes~\cite{Meng2005}.
Along this line, much work was done using either
relativistic~\cite{Guo2005,Fossion2006,Niksic2007,Li2009} or
nonrelativistic approaches~\cite{Rodriguez2007,Rodriguez2008,Robledo2008}.

Normally the regions of nuclei with strong octupole correlations
correspond to either the proton or neutron numbers close to 34
($1g_{9/2} \leftrightarrow 2p_{3/2}$ coupling), 56 ($1h_{11/2}
\leftrightarrow 2d_{5/2}$ coupling), 88 ($1i_{13/2} \leftrightarrow
2f_{7/2}$ coupling), and 134 ($1j_{15/2} \leftrightarrow 2g_{9/2}$
coupling)~\cite{Butler1996}. A variety of approaches were
applied to investigate the role of the octupole degree of freedom in Sm
and the neighboring nuclear region. The Woods-Saxon-Bogoliubov cranking
model is used to study the shapes of rotating Xe, Ba, Ce, Nd, and Sm
nuclei with $N=84-94$ and the expectations of octupole-deformed mean
fields at low and medium spins are confirmed~\cite{Nazarewicz1992}.
The {\it spdf} interacting boson model is applied to Sm isotopes
with $N=86-92$ to examine the signatures of octupole
correlations~\cite{Babilon2005}. Based on a collective
rotation-vibration Hamiltonian in which the axial quadrupole and
octupole degrees of freedom are coupled, the energy levels and
electromagnetic transition probabilities for $N=90$ isotones are
well reproduced~\cite{Minkov2006}.

Very recently, a new $K^\pi=0^-$ octupole excitation band was
observed in $^{152}$Sm and a pattern of repeating excitations built
on the $0_2^+$ level similar to those built on the ground state
emerges~\cite{Garrett2009}. It was suggested that $^{152}$Sm, rather
than a critical-point nucleus, is a complex example of shape
coexistence~\cite{Garrett2009}.

Based on the investigations mentioned previously, it is timely and
necessary to investigate the Sm isotopes in a microscopic and
self-consistent approach with the octupole degree of freedom.
The newly developed reflection-asymmetric relativistic mean-field
(RAS-RMF) approach is a good candidate for this
purpose~\cite{Geng2007} considering the remarkable success of RMF
theory~\cite{Ring1996,Vretenar2005,Meng2006} in describing many nuclear phenomena related to stable
nuclei~\cite{Ring1996}, exotic nuclei~\cite{Meng1996,Meng1998}, as
well as supernova and neutron stars~\cite{Glendenning2000}. In
Ref.~\cite{Geng2007}, the RAS-RMF approach was first applied
to the well-known octupole-deformed nucleus $^{226}$Ra, and
reproduced well both the binding energy and deformation.

In this article, the RAS-RMF approach will be applied to investigate
the potential energy surfaces (PES) of even-even $^{146-156}$Sm
isotopes in the ($\beta_2$,~$\beta_3$) plane and the shape
evolution involving the octupole degrees of freedom will be analyzed.

\section{Formalism}

The basic ansatz of the RMF theory is a Lagrangian density where
nucleons are described as Dirac particles that interact via the
exchange of various mesons and the photon. The mesons considered are
the isoscalar-scalar $\sigma$, the isoscalar-vector $\omega$, and the
isovector-vector $\rho$. The effective Lagrangian density
reads~\cite{Serot1986}

\begin{eqnarray} \nonumber
 \cal L&=&\bar\psi \left[
i\gamma^\mu\partial_\mu-M-g_\sigma\sigma-g_\omega\gamma^\mu\omega_\mu
-g_\rho\gamma^\mu \vec\tau \cdot \vec\rho_\mu - e\gamma^\mu\dfrac{1-\tau_3}{2}A_\mu \right] \psi\\
\nonumber
&&+\dfrac{1}{2}\partial^\mu\sigma\partial_\mu\sigma-\dfrac{1}{2} m_\sigma^2\sigma^2
-\dfrac{1}{3}g_2\sigma^3-\dfrac{1}{4}g_3\sigma^4\\
\nonumber
&&-\dfrac{1}{4}\Omega^{\mu\nu}\Omega_{\mu\nu}+\dfrac{1}{2}m_\omega^2\omega^\mu\omega_\mu
+\dfrac{1}{4}c_3 (\omega^\mu\omega_\mu)^2\\
\nonumber
&&-\dfrac{1}{4}\vec R^{\mu\nu}\cdot\vec R_{\mu\nu}+\dfrac{1}{2}m_\rho^2\vec\rho^\mu\cdot\vec\rho_\mu\\
&&-\dfrac{1}{4}F^{\mu\nu}F_{\mu\nu},
 \label{Lagrangian}
\end{eqnarray}
in which the field tensors for the vector mesons and the photon are,
respectively, defined as
\begin{eqnarray}
\left\{
\begin{array}{lll}
   \Omega_{\mu\nu}   &=& \partial_\mu\omega_\nu-\partial_\nu\omega_\mu, \\
   {\vec R}_{\mu\nu} &=& \partial_\mu{\vec \rho}_\nu
                        -\partial_\nu{\vec \rho}_\mu, \\
   F_{\mu\nu}        &=& \partial_\mu A_\nu-\partial_\nu  A_\mu.
\end{array}   \right.
\end{eqnarray}

Using the classical variational principle, one can obtain
the Dirac equation for the nucleons and the Klein-Gordon
equations for the mesons. To solve these
equations, we employ the basis expansion method,
which was widely used in both the nonrelativistic
and relativistic mean-field models. For axial-symmetric
reflection-asymmetric systems,
where nonaxial quadrupole and octupole deformations are plainly excluded,
the spinors are expanded in terms of the eigenfunctions of the
two-center harmonic-oscillator (TCHO) potential
\begin{eqnarray}
V(r_\perp,z)= \dfrac{1}{2} M \omega_\perp^2 r_\perp^2 +\left\{
\begin{array}{ll}
   \dfrac{1}{2} M \omega_1^2(z+z_1)^2, & z<0\\
   &\\
   \dfrac{1}{2} M \omega_2^2(z-z_2)^2, & z\geqslant0\\
\end{array}   \right.
\end{eqnarray}
where $M$ is the nucleon mass, $z_1$ and $z_2$ (real, positive)
represent the distances between the centers of
the spheroids and their intersection planes, and $\omega_1$($\omega_2$)
are the corresponding oscillator frequencies for $z < 0$ ($z \geqslant 0$)~\cite{Geng2007}.
The TCHO basis was widely used in studies of fission, fusion,
heavy-ion emission, and various cluster phenomena~\cite{Greiner1994}.
By setting proper asymmetric parameters,
the major and the $z$-axis quantum numbers are real numbers very close to integers,
and the integers are used in the Nilsson-like notation $\Omega$[$N$$n_z$$m_l$] for convenience.
More details can be found in Ref.~\cite{Geng2007}.

The binding energy at a certain deformation is obtained by
constraining the mass quadrupole moment $\langle \hat{Q_2}\rangle $
to a given value $\mu_2$~\cite{Ring1980}
 \beq
  \langle H'\rangle ~=~\langle H\rangle  +  \dfrac{1}{2}C(\langle \hat{Q_2}\rangle -\mu_2)^2
 \eeq
where $C$ is the curvature constant parameter and $\mu_2$ is
the given quadrupole moment. The expectation value of $\hat{Q_2}$ is
 $\langle \hat{Q_2}\rangle =\langle \hat{Q_2}\rangle _n+\langle \hat{Q_2}\rangle
 _p$ with
 $\langle\hat{Q_2}\rangle _{n,p}= \langle 2 r^2 P_2(\cos\theta)\rangle _{n,p}$.
The deformation parameter $\beta_2$ is related to $\langle
\hat{Q_2}\rangle $ by $\langle \hat{Q_2}\rangle  =
\dfrac{3}{\sqrt{5\pi}} Ar^2\beta_2$ with $r = R_0 A^{1/3}$
($R_0=1.2$ fm) and $A$ the mass number. The octupole moment
constraint can also be applied similarly with
$\langle\hat{Q_3}\rangle =\langle \hat{Q_3}\rangle _n+\langle
\hat{Q_3}\rangle _p$, $\langle\hat{Q_3}\rangle _{n,p}= \langle 2 r^3
P_3(\cos\theta)\rangle _{n,p}$, and $\langle\hat{Q_3}\rangle  =
\dfrac{3}{\sqrt{7\pi}} Ar^3\beta_3$. By constraining the quadruple
moment and octupole moment simultaneously, the total energy surface
in the ($\beta_2$,~$\beta_3$) plane can be obtained.

\section{Results and discussion}

The properties of even-even $^{146-156}$Sm are calculated
in the constrained RAS-RMF approach with parameter set PK1~\cite{Long2004}.
The parameter set PK1 is one of the best parameter sets available
in the framework of RMF theory, which, as usual,
is obtained by fitting the masses of selected spherical nuclei
as well as saturation properties of nuclear matter.
The success of universal RMF parameter sets was demonstrated for
describing the properties of spherical~\cite{Ring1996} and deformed nuclei~\cite{Vretenar2005,Meng2006},
and they are believed to be appropriate for application in octupole-deformed nuclei.
In return, the application for octupole-deformed nuclei will also provide a further test.
The TCHO basis with 16 major shells for both fermions and bosons is used.
The pairing correlations are treated by the BCS approximation with
a constant pairing gap $\Delta=11.2/\sqrt{A}$ MeV.

The binding energy, quadrupole, and octupole deformations
are listed for the ground states of $^{146-156}$Sm in Table~\ref{tab:be}.
The binding energies are well reproduced within 0.2\%. Moreover,
excellent agreement is obtained for the quadrupole deformations
except $^{152}$Sm (see the discussion in the following).

\begin{table}\tabcolsep=4pt
\caption{
The total binding energy (in MeV) as well as the quadrupole deformation $\beta_2$
and octupole deformation $\beta_3$ of the ground states of even-even $^{146-156}$Sm obtained in
the constrained RAS-RMF approach with PK1,
in comparison with the available experimental data.}
\small
\begin{center}
\begin{tabular}{c|ccc|cc}
\hline\hline
Nucleus & $E^{\rm cal}$ &$\beta_2^{\rm cal}$ &$\beta_3^{\rm cal}$&
$E^{\rm exp}$~\cite{Audi2003} &$\beta_2^{\rm exp}$~\cite{Raman2001} \\
\hline
$^{146}$Sm&  1213.38& 0.09& 0.07& 1210.91&  $-$ \\
$^{148}$Sm&  1227.27& 0.14& 0.08& 1225.39&  0.14\\
$^{150}$Sm&  1241.59& 0.18& 0.13& 1239.25&  0.19\\
$^{152}$Sm&  1254.46& 0.20& 0.15& 1253.10&  0.31\\
$^{154}$Sm&  1267.76& 0.32& 0.00& 1266.94&  0.34\\
$^{156}$Sm&  1280.08& 0.33& 0.02& 1279.99&  $-$ \\

\hline\hline
\end{tabular}
\end{center}
\label{tab:be}
\end{table}

\begin{figure}[htb]
\includegraphics[scale=0.36]{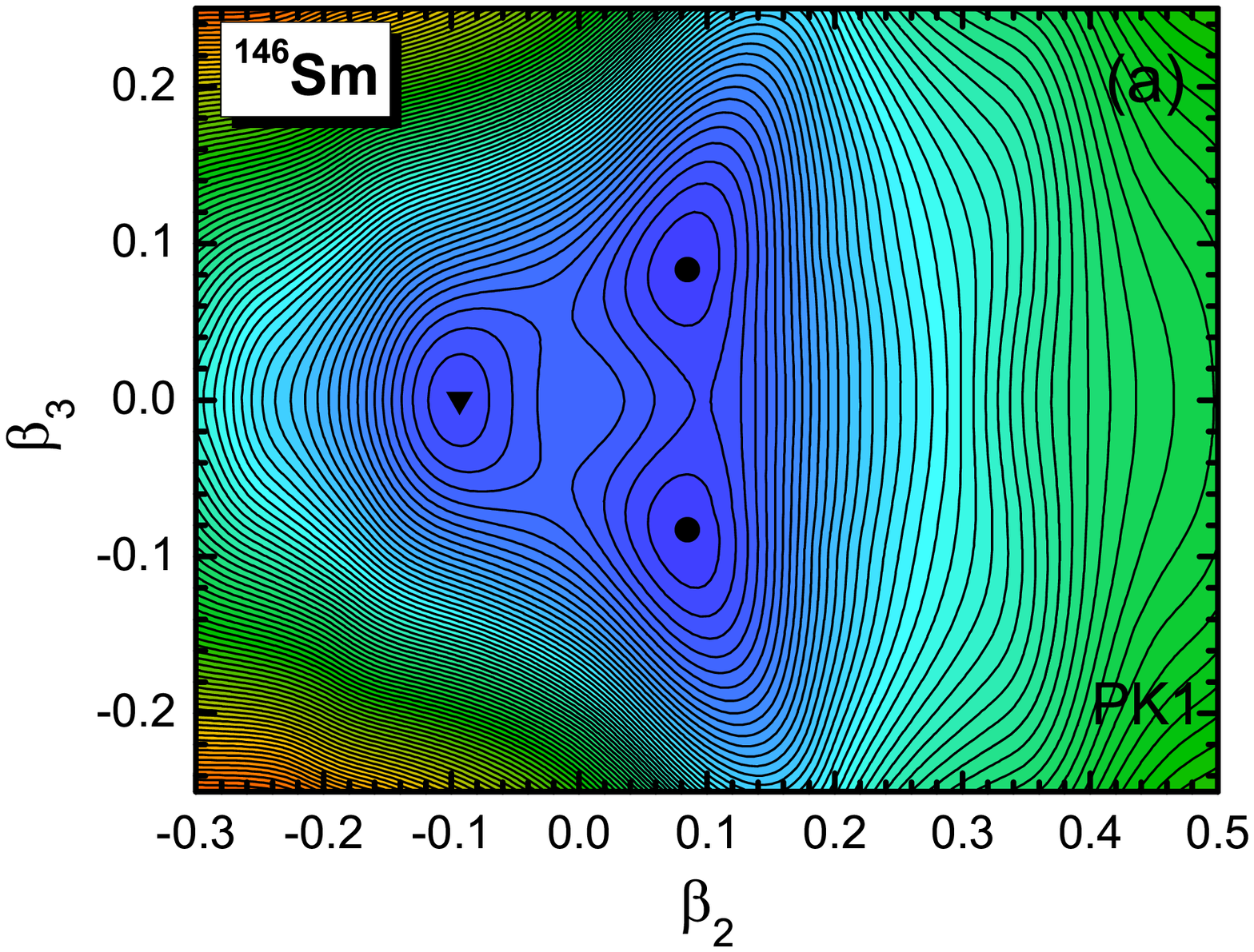}
\includegraphics[scale=0.36]{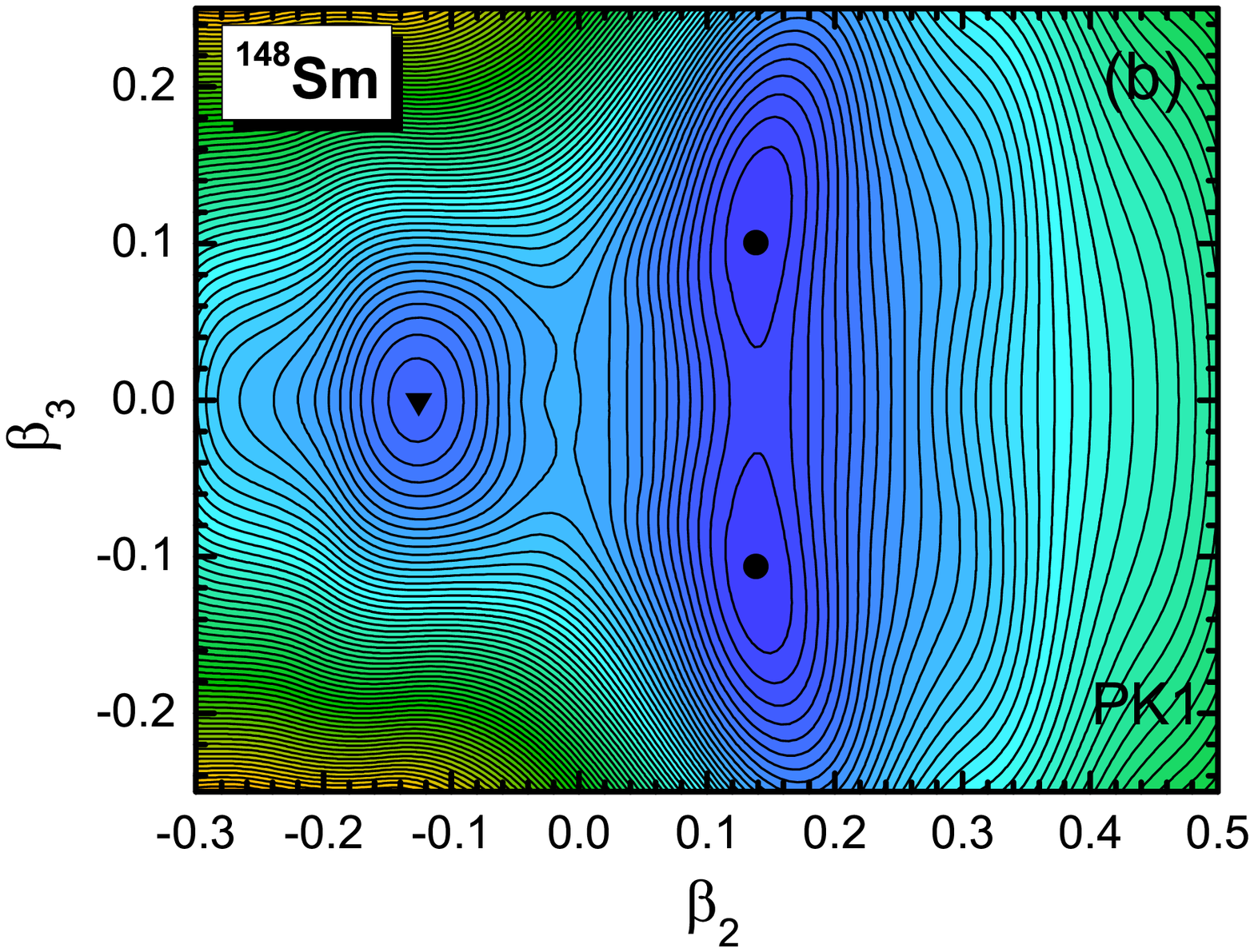}\\
\includegraphics[scale=0.36]{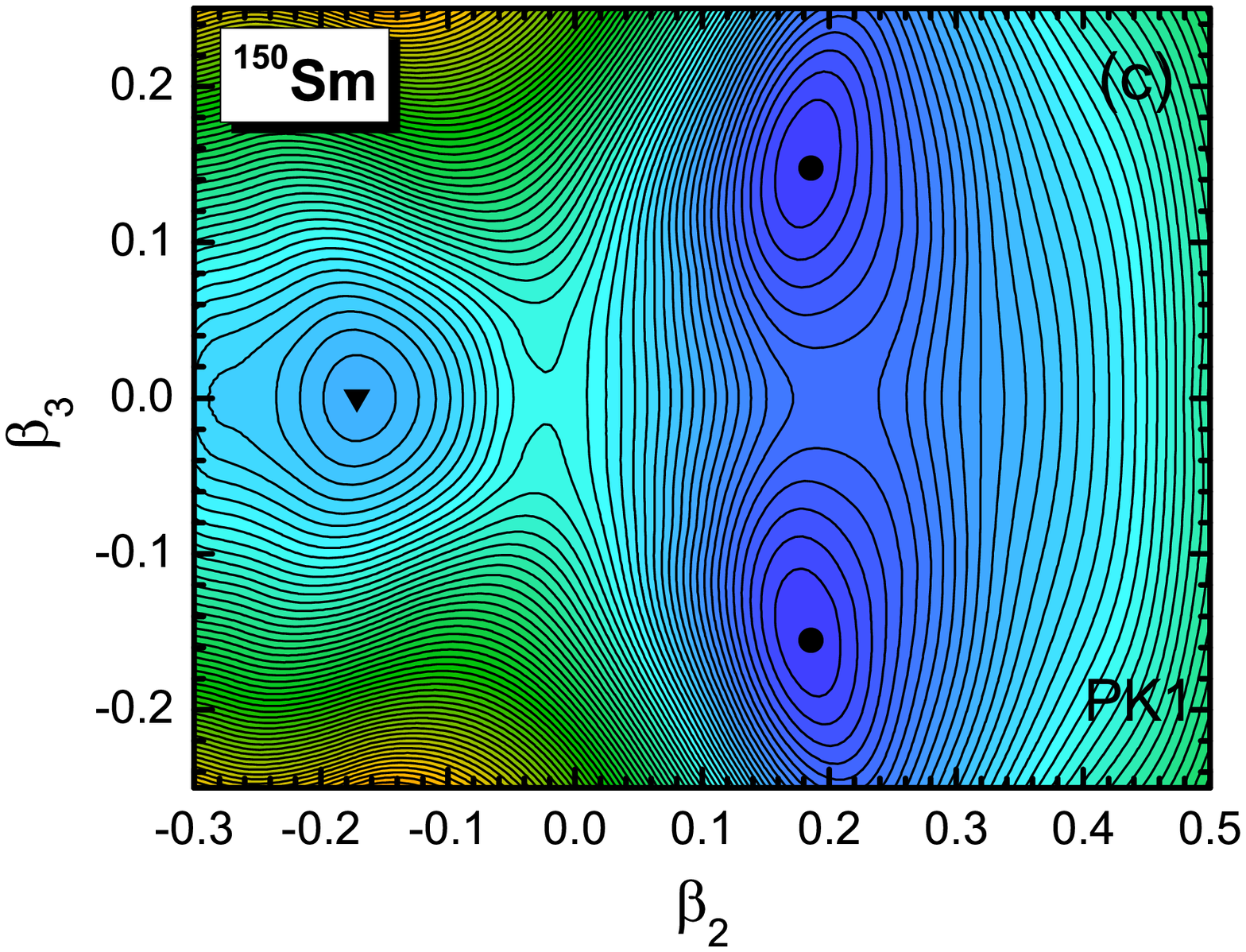}
\includegraphics[scale=0.36]{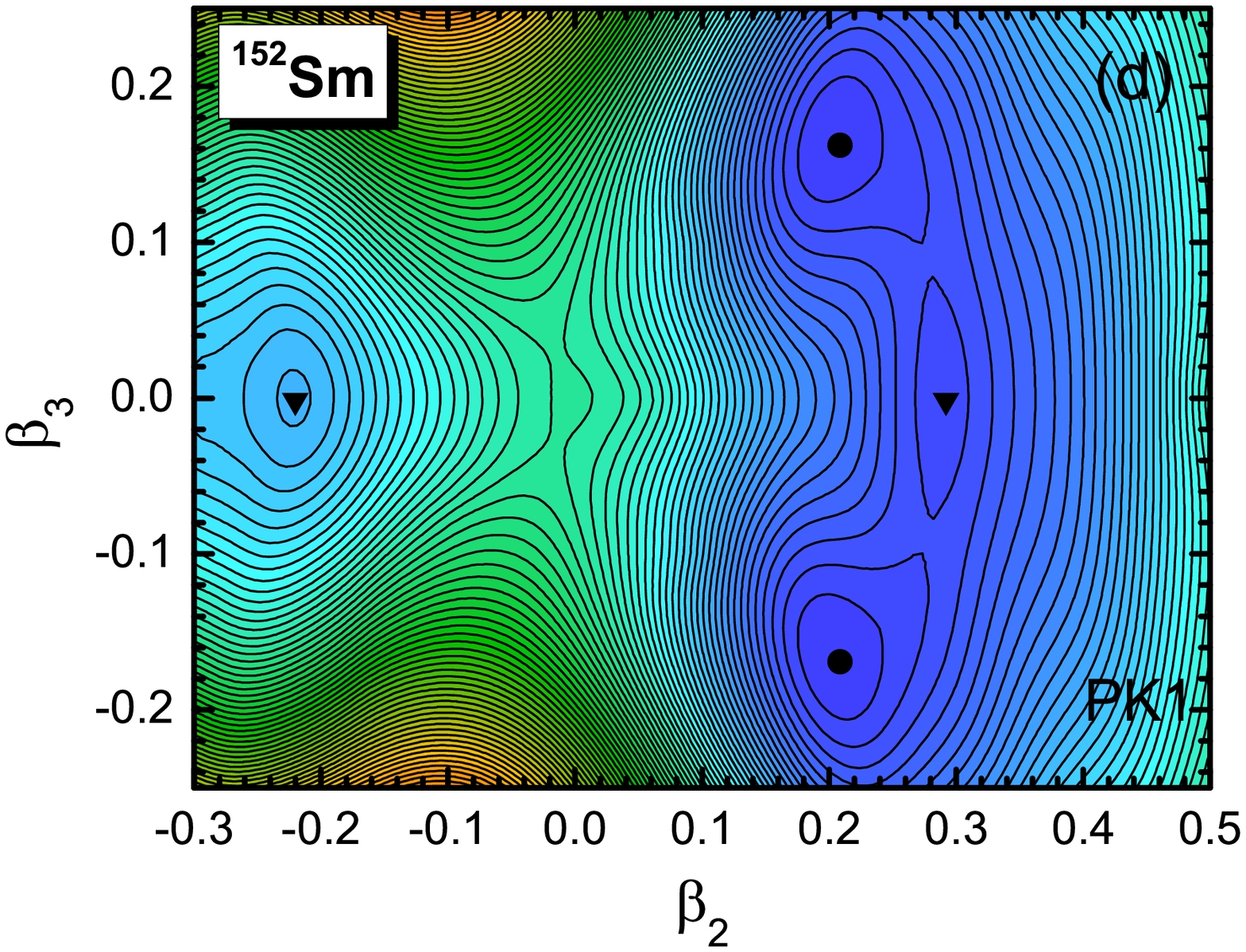}\\
\includegraphics[scale=0.36]{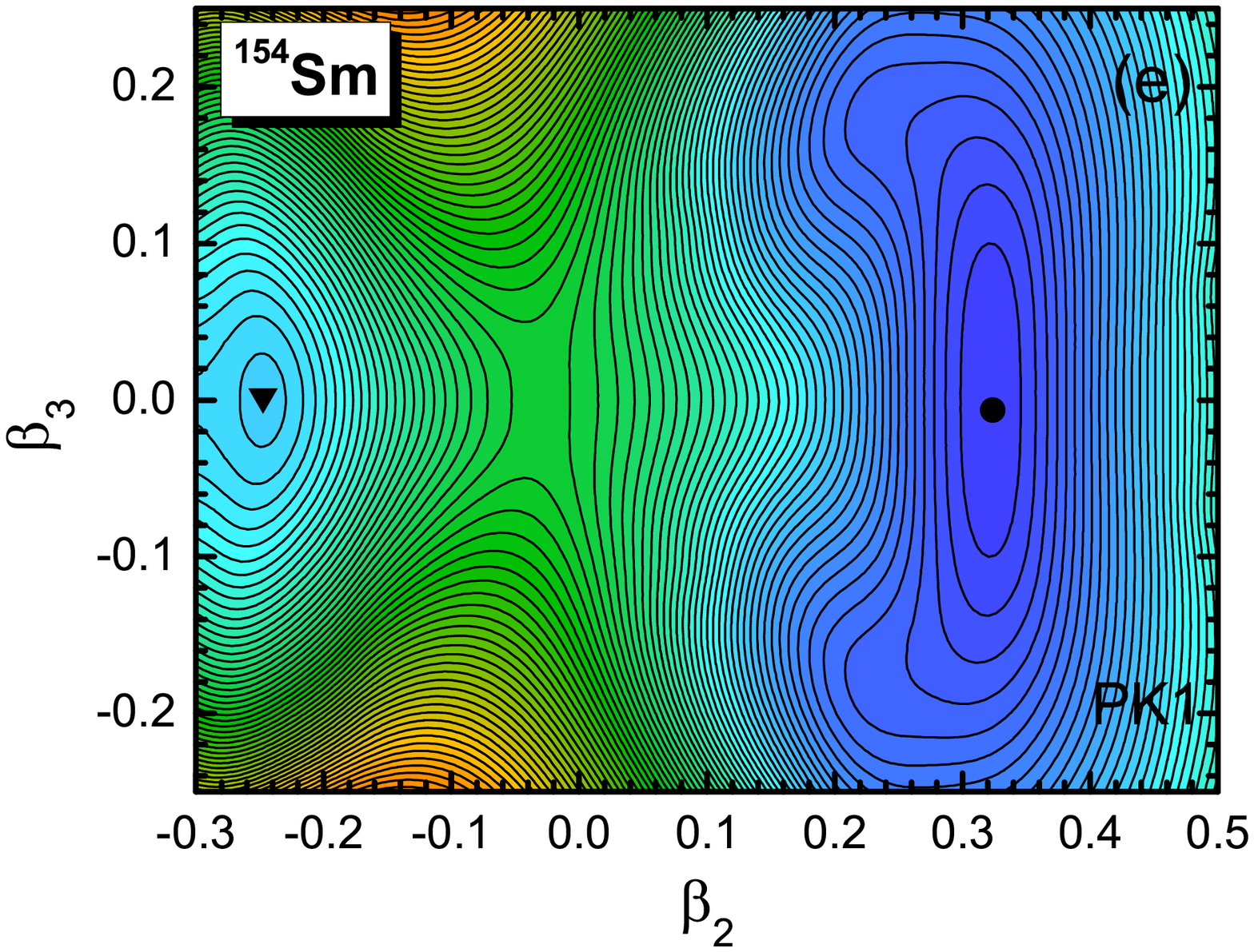}
\includegraphics[scale=0.36]{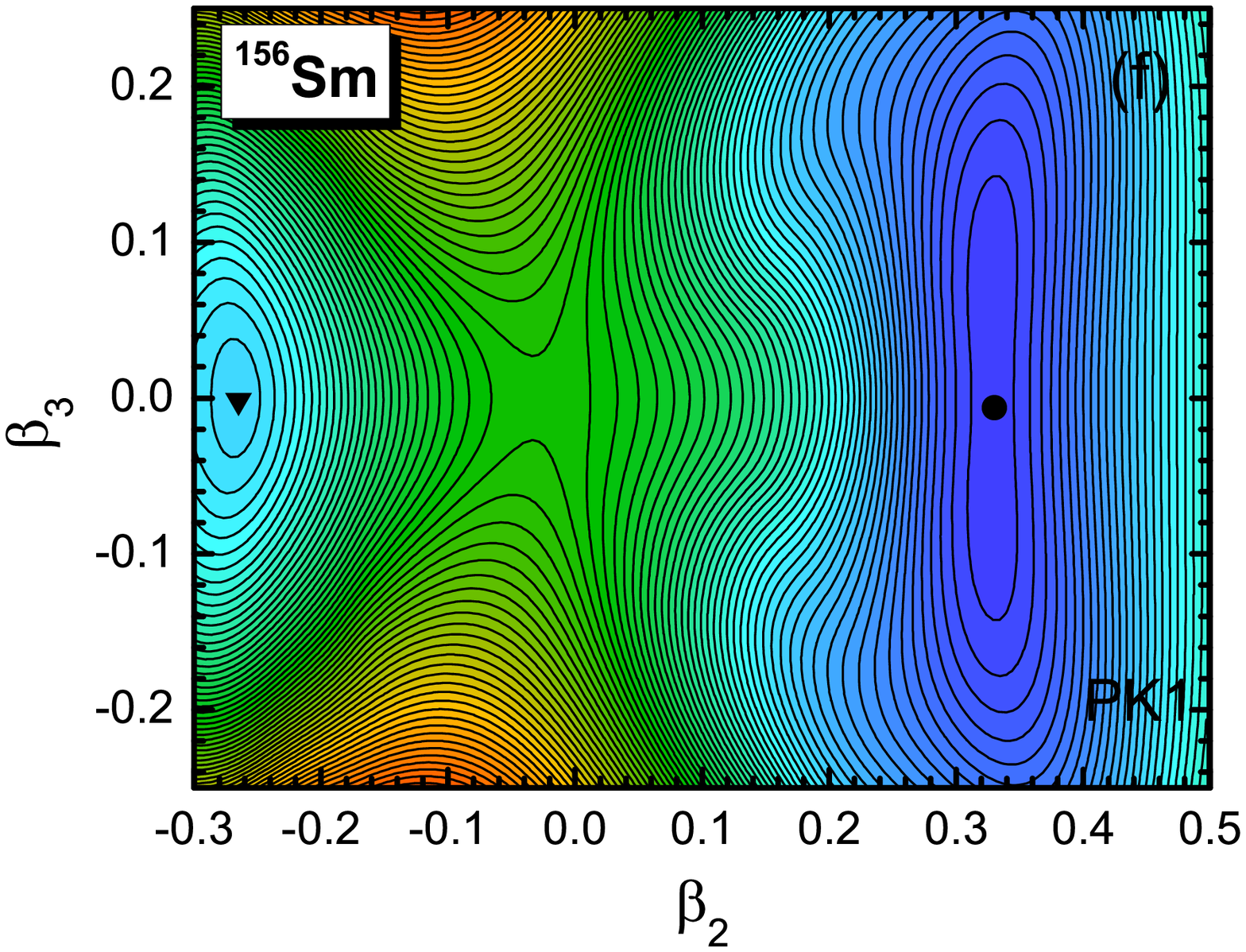}\\
\caption{(Color online)
The contour plots of total energies for the even-even $^{146-156}$Sm in ($\beta_2$,~$\beta_3$) plane
obtained in the RAS-RMF approach with PK1 and constant-$\Delta$ pairing.
The energy separation between contour lines is 0.25 MeV.
The global minima and other local minima are denoted by ``$\bullet$" and ``$\blacktriangledown$", respectively.
}
\label{fig:pes}
\end{figure}

To investigate the shape evolution in the Sm isotopes,
the contour plots of the total energies as functions of $\beta_2$ and $\beta_3$
for $^{146-156}$Sm are shown in Fig.~\ref{fig:pes},
which have up-down symmetry in the ($\beta_2$,~$\beta_3$) plane
because of the equivalence between the states with positive and negative $\beta_3$.

It is shown in Fig.~\ref{fig:pes} that for the ground states,
$^{146,148}$Sm are near spherical,
$^{150}$Sm octupole deformed, and $^{154,156}$Sm well deformed
while $^{152}$Sm marks the transition from octupole to quadrupole deformed.
In detail, for $^{146}$Sm, with $N=84$ close to the magic number 82,
the ground state is near spherical with ($\beta_2,~\beta_3$)=(0.08, 0.08).
For $^{148}$Sm and $^{150}$Sm, with increasing neutron number,
the deformations ($\beta_2,~\beta_3$) gradually increase.
Particularly for $^{150}$Sm with $N=88$, a global minimum with
substantial quadrupole and octupole deformations($\beta_2,~\beta_3$)=(0.19, 0.14)
is well developed, which is about 1.36 MeV deeper than
the corresponding quadrupole-deformed state.
For $^{152}$Sm, the global minimum moves to ($\beta_2,~\beta_3$)=(0.20, 0.15),
while a quadrupole minimum emerges at ($\beta_2,~\beta_3$)=(0.29, 0).
It is interesting to note that the deformation of this minimum is quite close to the
experimental value $\beta_2^{\rm exp} = 0.31$~\cite{Raman2001}
listed in Table~\ref{tab:be}. The energy difference between the two
minima is 0.33 MeV with a 0.5 MeV barrier in between. For
$^{154,156}$Sm, the ground states are well quadrupole deformed with
($\beta_2$,~$\beta_3$) $\sim$ (0.33, 0).
Similar PES's can also be obtained with other parameter sets such as NL3~\cite{Lalazissis1997}.
In addition, one notes that
the oblate minima shown in Fig.~\ref{fig:pes} may not be stable
against the $\gamma$ direction~\cite{Li2009}.

In Ref.~\cite{Meng2005}, axially deformed RMF calculation with a
variety of effective interactions was performed for
$^{144-158}$Sm to discuss the transition from spherical $U(5)$ to
axially deformed $SU(3)$ shapes. It was shown that the PES's of
$^{144,146}$Sm are minimized near spherical and of $^{154-158}$Sm
well-deformed, while in between the PES's of $^{148,150,152}$Sm are
found to be relatively flat. From Fig.~\ref{fig:pes}, it is shown
that the conclusion remains to be true even with the inclusion of
the octupole degree of freedom.

Meanwhile, we do find something new for $^{152}$Sm after including
the octupole degree of freedom, that is, $^{152}$Sm marks the
shape/phase transition not only from $U(5)$ to $SU(3)$ symmetry, but
also from the octupole-deformed to quadrupole-deformed case.

Quite recently in Ref.~\cite{Garrett2009}, a new $K^\pi=0^-$ band
was observed, which has a remarkable similarity in its $E1$
transition to the first excited $K^\pi=0^+$  band as the lowest
$K^\pi=0^-$ band to the ground-state band. A pattern of repeating
excitations built on the $0^+_2$  level similar to those built
on the ground state is claimed to indicate that $^{152}$Sm is a
complex example of shape coexistence rather than a critical-point
nucleus. These observations can be well understood from the PES
obtained previously. For $^{152}$Sm with the octupole minimum at
($\beta_2,~\beta_3$)=(0.20, 0.15) and the quadrupole minimum at
($\beta_2,~\beta_3$)=(0.29, 0), if one performs the generator
coordinate method (GCM)~\cite{Ring1980,Yao2009} calculation with the
PES, two low-lying states in the ($\beta_2$,$\beta_3$) plane with
similar quadrupole deformation will be obtained, which are a mixture
of quadrupole and octupole deformation configurations. Based on
these two states, the pattern of repeating excitations is expected.

\begin{figure}[!htb]
\includegraphics[scale=0.58]{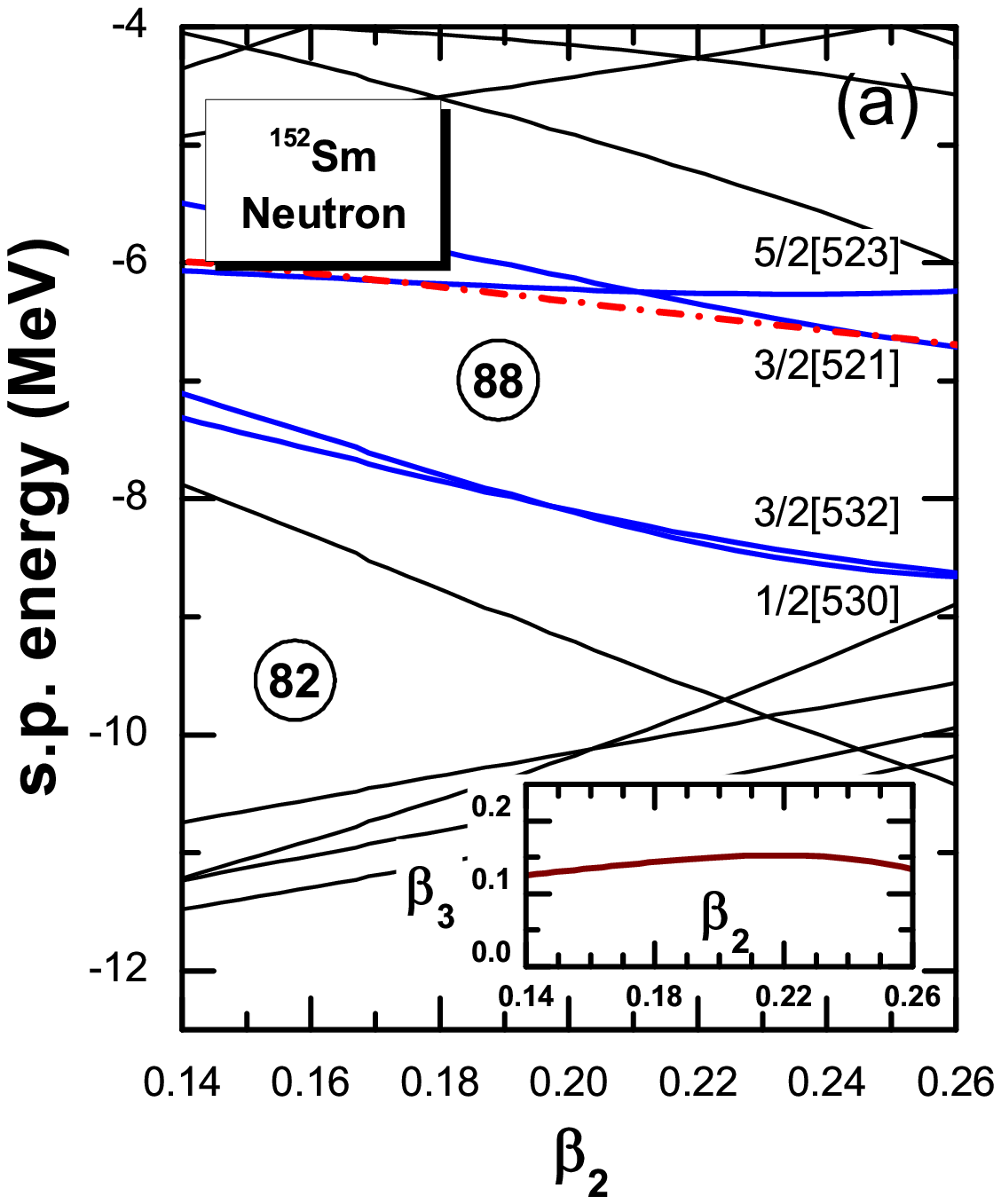}
\hspace{0.5cm}
\includegraphics[scale=0.58]{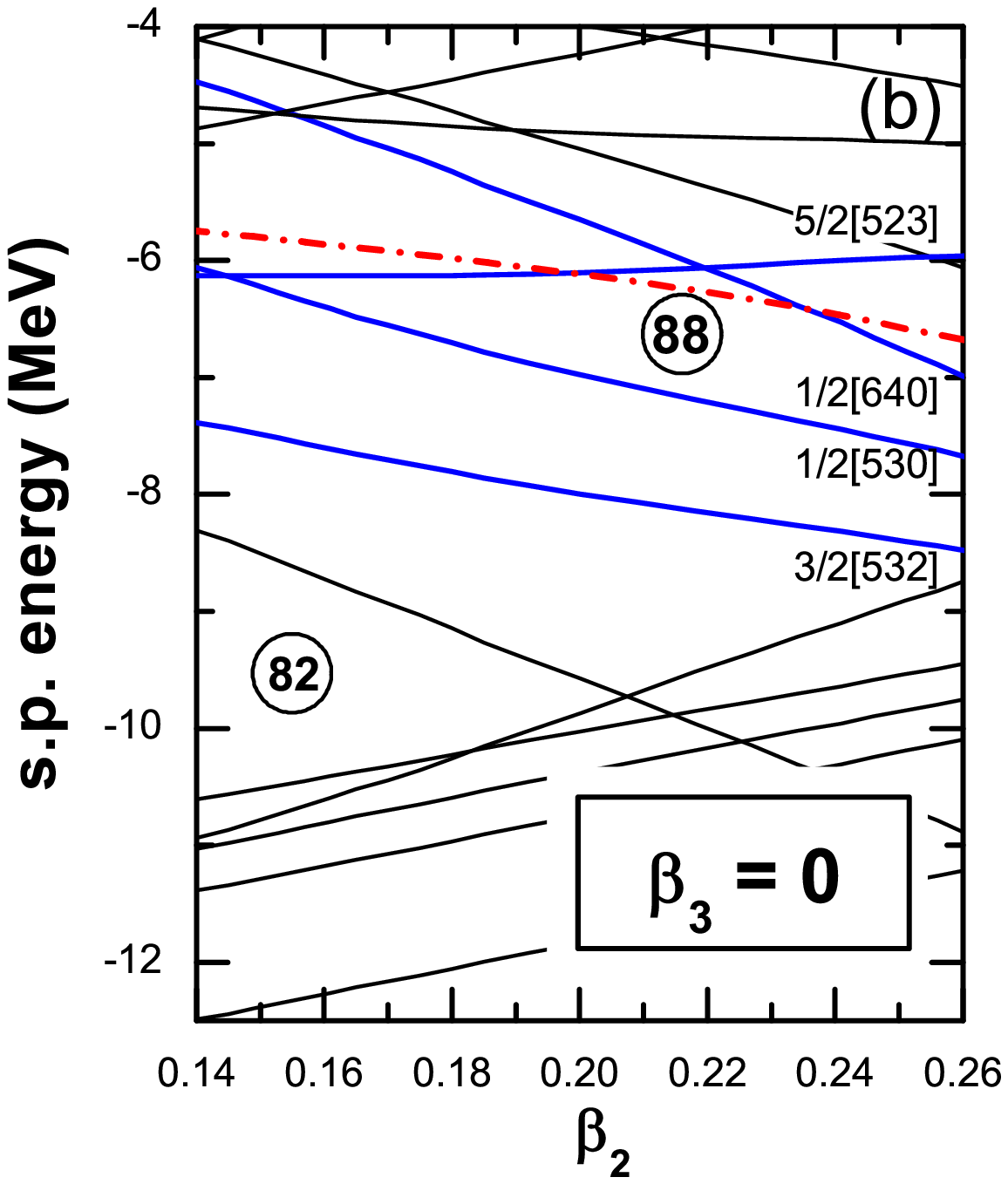}
\caption{(Color online)
Neutron single-particle levels of $^{152}$Sm in RAS-RMF approach
with PK1 as functions of $\beta_2$ for
states minimized with respect to $\beta_3$ (left panel) and states with $\beta_3 =0$ (right panel).
The dash-dot lines denote the corresponding Fermi surfaces.
The levels near the Fermi surface are labeled by
Nilsson-like notations $\Omega$[$N$$n_z$$m_l$] of the first component at $\beta_2=0.20$.
The corresponding $\beta_3$ are shown in the inset.
}
\label{fig:sp1}
\end{figure}

\begin{figure}[!htb]
\includegraphics[scale=0.58]{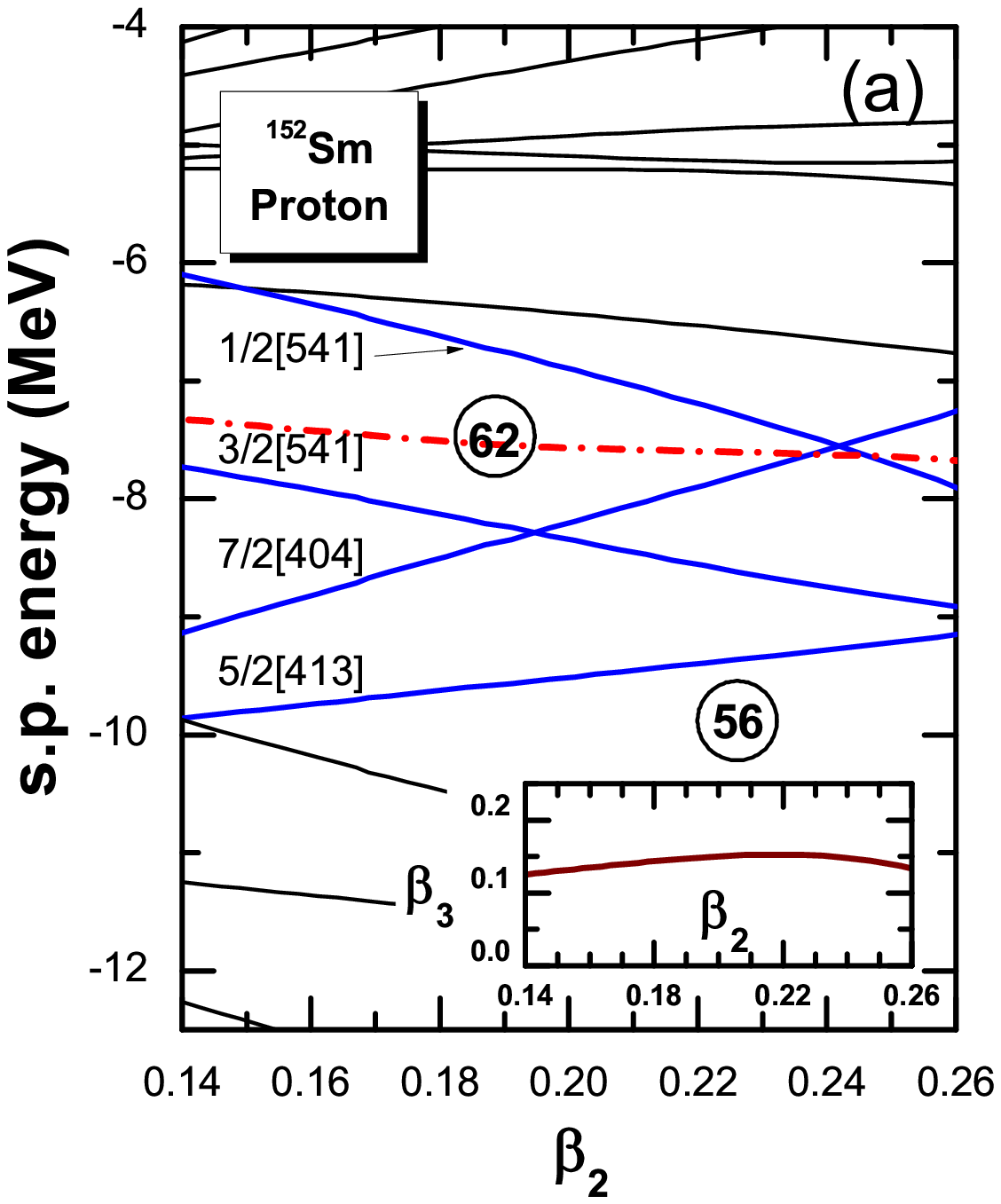}
\hspace{0.5cm}
\includegraphics[scale=0.58]{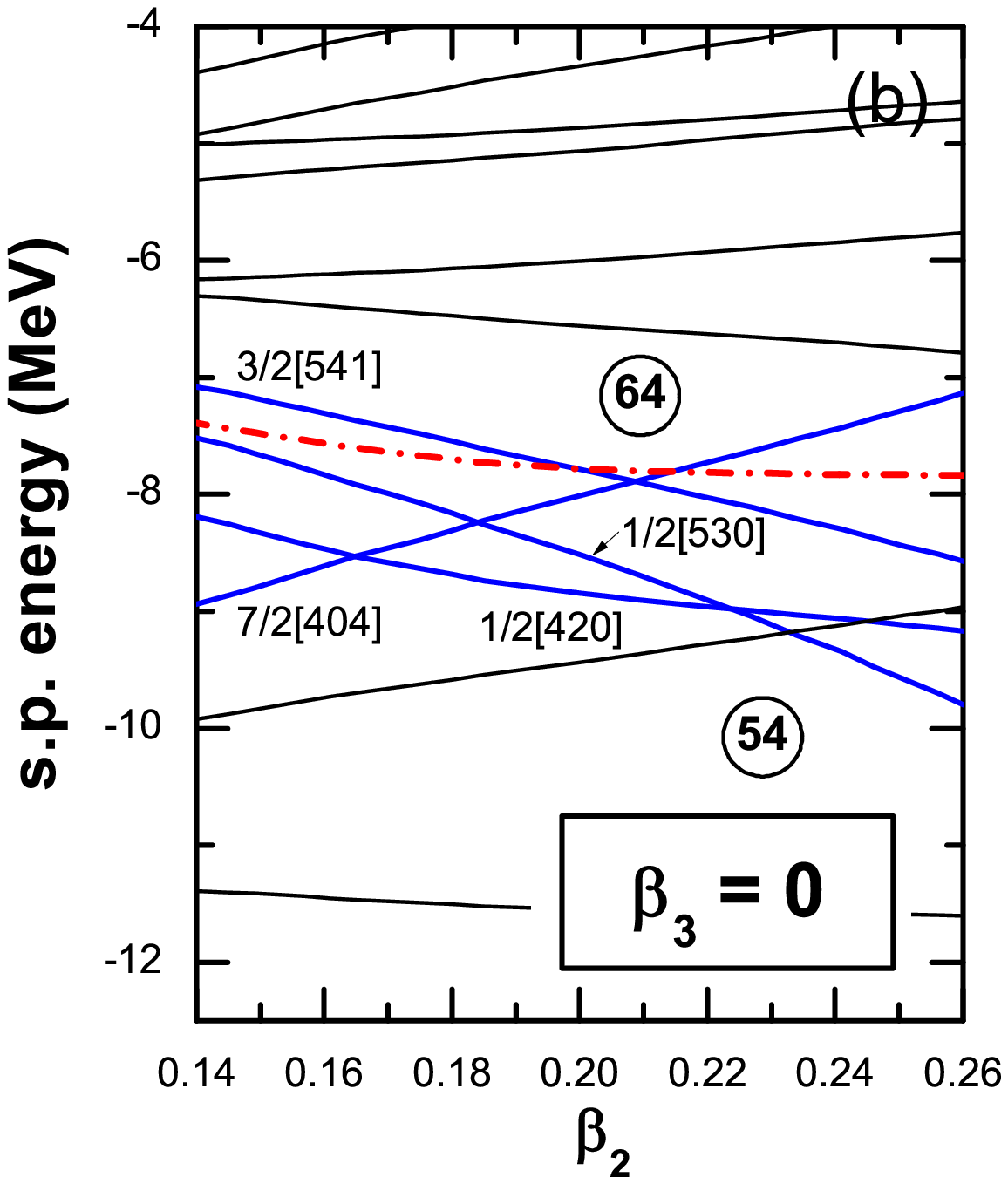}
\caption{(Color online)
Same as Fig.~\ref{fig:sp1}, but for proton.}
\label{fig:sp2}
\end{figure}

To understand the evolution of the octupole deformation
microscopically, the neutron single-particle levels in $^{152}$Sm
for the states minimized with respect to $\beta_3$ and the states
with $\beta_3=0$ for $\beta_2=0.14 \sim 0.26$ are shown in Fig.
~\ref{fig:sp1}. The levels near the Fermi surface are labeled by
Nilsson-like notations $\Omega[Nn_zm_l]$ of the largest component.
In the left panel of Fig.~\ref{fig:sp1}, a large energy gap with
$N=88$ near the Fermi surface is found, which is related to the
softness of the potential energy surface in the quadrupole and octupole
degrees of freedom in $^{152}$Sm. There is no obvious neutron gap
near the Fermi surface for the states with $\beta_3=0$.
In addition, the proton single-particle levels for the corresponding states
are shown in Fig.~\ref{fig:sp2}, where no obvious
energy gaps can be found near the Fermi surfaces.

It is well known that for nuclei with $N\sim88$ or $Z\sim56$ the
octupole deformation driving pairs of orbitals include ($\nu
2f_{7/2}$, $\nu 1i_{13/2}$) and ($\pi 2d_{5/2}$, $\pi 1h_{11/2}$),
which in the axially deformed case will be subgrouped as
($\nu1/2[541]$, $\nu1/2[660]$), ($\nu3/2[532]$, $\nu3/2[651]$),
($\nu5/2[523]$,$\nu5/2[642]$), ($\nu7/2[514]$, $\nu7/2[633]$), and
($\pi1/2[431]$, $\pi1/2[550]$), ($\pi3/2[422]$, $\pi3/2[541]$),
($\pi5/2[413]$, $\pi5/2[532]$), respectively. It is interesting to
investigate the performance of such pairs in the single-particle
levels near the Fermi surfaces in Figs.~\ref{fig:sp1} and \ref{fig:sp2}.
These levels together with their BCS occupation probabilities and
corresponding contributions from the four leading components are
shown in Table~\ref{tab:com}. Taking the level $\nu 3/2[521]$ as an
example, its second (20.1\%) and third (15.5\%) components compose
an octupole deformation driving pair ($\nu3/2[532]$, $\nu3/2[651]$).
Similarly, one can find
the pair ($5/2[523]$, $5/2[642]$) for $\nu 5/2[523]$, the pair
($3/2[532]$, $3/2[651]$) for $\nu 3/2[532]$, and the pair
($1/2[541]$, $1/2[660]$) (the fifth component $1/2[660]$ with 6.0\%,
which is not listed in Table~\ref{tab:com}) for $\nu 1/2[530]$.
However, for the proton side, no octupole deformation driving pairs are found
among the four leading components.
Therefore the neutron orbital takes a more important role in the
evolution of octupole deformation  in Sm isotopes than the proton
one, consistent with the energy gaps presented in Fig.~\ref{fig:sp1}.

\begin{table}\tabcolsep=4pt
\caption{Single-particle levels near the Fermi surface for the ground
state with ($\beta_2,~\beta_3$)= (0.20, 015) in $^{152}$Sm together with their BCS occupation probabilities
and corresponding contributions from the four leading components.
The components originating from the octupole deformation driving pairs of orbitals
($\nu 2f_{7/2}$, $\nu 1i_{13/2}$) and ($\pi 2d_{5/2}$, $\pi 1h_{11/2}$) are in bold.
}
\small
\begin{center}
\begin{tabular}{cccccccccc}
\hline\hline
level  & occu. & \multicolumn{2}{c}{1st comp.} & \multicolumn{2}{c}{2nd comp.}  &
\multicolumn{2}{c}{3rd comp.} &\multicolumn{2}{c}{4th comp.} \\ \hline
$\nu 3/2[521]$& 0.401&
 3/2[521]& 33.3\% &
{\bf 3/2[532]}&{\bf 20.1\%} &
{\bf 3/2[651]}&{\bf 15.5\%} &
 3/2[631]&  9.4\% \\
$\nu 5/2[523]$& 0.434&
{\bf 5/2[523]}&{\bf 57.2\%} &
 5/2[532]& 16.0\% &
{\bf 5/2[642]}&{\bf  6.7\%} &
 5/2[633]&  5.2\% \\
$\nu 3/2[532]$& 0.946&
{\bf 3/2[532]}&{\bf 46.4\%} &
 3/2[541]& 21.1\% &
 3/2[512]&  8.8\% &
{\bf 3/2[651]}&{\bf  5.8\%} \\
$\nu 1/2[530]$& 0.947&
 1/2[530]& 33.9\% &
{\bf 1/2[541]}&{\bf 18.1\%} &
 1/2[510]&  8.0\% &
 1/2[651]&  6.2\% \\
\hline
$\pi 1/2[541]$& 0.219&
 1/2[420]& 21.4\% &
 1/2[541]& 21.4\% &
 1/2[440]& 17.7\% &
 1/2[521]&  9.8\% \\
$\pi 7/2[404]$& 0.763&
 7/2[404]& 85.9\% &
 7/2[413]&  7.3\% &
 7/2[514]&  2.8\% &
 7/2[604]&  0.4\% \\
$\pi 3/2[541]$& 0.834&
{\bf 3/2[541]}&{\bf 34.6\%} &
 3/2[411]& 20.0\% &
 3/2[521]& 17.9\% &
 3/2[532]&  9.1\% \\
$\pi 5/2[413]$& 0.952&
{\bf 5/2[413]}&{\bf 76.0\%} &
 5/2[422]& 11.2\% &
 5/2[523]&  4.7\% &
 5/2[303]&  2.0\% \\
\hline\hline
\end{tabular}
\end{center}
\label{tab:com}
\end{table}

\section{Conclusion}
In this article, the PES's of even-even $^{146-156}$Sm in the
($\beta_2,~\beta_3$) plane are obtained by the constrained RAS-RMF
approach, and the single-particle levels near the Fermi surfaces for the
nucleus $^{152}$Sm are studied. It is shown that the critical-point
candidate nucleus $^{152}$Sm marks the shape/phase transition not
only from $U(5)$ to $SU(3)$ symmetry, but also from the octupole-deformed 
ground state in $^{150}$Sm  to quadrupole-deformed ground
state in $^{154}$Sm.

Furthermore, the microscopic PES for the nucleus $^{152}$Sm is
consistent with the claimed shape coexistence based on the
observation of repeating excitations built on the $0^+$ level
similar to those built on the ground state~\cite{Garrett2009}.

By including the octupole degree of freedom, an energy gap near
the Fermi surface for single-particle levels in $^{152}$Sm with $N=88$
and $\beta_2 = 0.14 \sim 0.26$ is found, which is related to the
softness of its potential energy surface in the quadrupole and octupole
degrees of freedom. From the energy gap and the components of the
single-particle levels near the Fermi surface, it is demonstrated that
the neutrons play an important role for the octupole deformation
driving in $^{152}$Sm.

\section*{ACKNOWLEDGMENTS}
This work was supported, in part, by the
Major State Basic Research Developing Program 2007CB815000,
National Natural Science Foundation of China under Grant Nos 10775004, 10975007, and 10975008,
China Postdoctoral Science Foundation,
the Natural Science Foundation of He'nan Educational Committee under Grant No. 200614003, and
the Young Backbone Teacher Support Program of He'nan Polytechnic University.

%\end{CJK*}

\end{document}